# Quality of Service (QoS): Measurements of Video Streaming

**Sajida Karim[1], Hui He[1,*] , Asif Ali Laghari[1] and Hina Madiha[2]**

**[1] School of Computer Science & Technology, Harbin Institute of Technology, Harbin, China**

**[2] School of Computer Science &Technology, Shaanxi Normal University, Xian, China**

**Abstract**

Nowadays video streaming is growing over the social clouds, where end-users always want to share High Definition (HD) videos among friends. Mostly videos were recorded via smartphones and other HD devices and short time videos have a big file size. The big file size of videos required high bandwidth to upload and download on the Internet and also required more time to load in a web page for play. So avoiding this problem social cloud compress videos during the upload for smooth play and fast loading in a web page. Compression decreases the video quality which also decreases the quality of experience of end users. In this paper we measure the QoS of different standard video file formats on social clouds; they varied from each other in resolution, audio/video bitrate, and storage size. Videos were downloaded from YouTube and converted standard video file formats which provided by developers and upload/download at social clouds. During the upload of video standard file format, original video format converted in social cloud's default supported file format and also compressed according to compression technique which set by the service provider. The Result shows that Facebook compressed HD Videos more as compared to Google+ videos. However, Facebook gives a better quality of compressed videos on slow networks, which were upload in the format of MOV, FLV, WMV, AVI, WebM and converted by Facebook into MP4 format for low resolutions and MP4 (HEVC and AVC) for 1080P and 2160P high resolutions. Google+ support all file formats and did not convert file format and provide high quality video streaming for high speed networks.

## 1. Introduction

Many video file formats (codecs) were developed by the different organization to provide better visual quality with low file size which saves storage and easily stream on the slow network without buffer/delay. Social clouds are attractive for users to sharing and watch their personal and other information videos among the community. The user has different devices such as smartphones, DSLR camera, movie recorder to record adventure and family parties' videos and share on social media [1]. Heterogeneous devices use different file format to store recorded videos and they were upload on social media by using mobile apps of social media service providers [2, 3].

Different standard video file formats codec plays a significant role in providing video streaming for a low bit and frame rates while preserving the high-quality services so that they maximized networks, reduced file size and Compression parameters for providing better services and video quality [4]. Compression mostly works on the lossy technique that compressed lacks of some information present in the original videos. Concerning the video streaming, the video stream is compressed using a video codec such as H.264 or VP8. Encoded video streams assemble in the container of bit stream of video file formats such in MP4, FLV, WEBM and others video file formats. A variety of video compression formats can be implemented with multiple codecs in the same file; many video codecs use common to make them compatible and standard video compression formats [5].

Flash videos were used for web-based streaming via real time messaging protocol (RTMP), where Adobe flash player integrated into the browser was used to watch videos before 2000. [6]. Flash video support two file formats with extension. FLV and .F4V. Flash video FLV files encode the Sorenson Spark or VP6 codec and video compression .

In 1998, MOV- QuickTime Movie was developed by Apple Inc. and Mov files use the MPEG-4 codec for compression. The ISO approved the MPEG-4 file formats that support MP4 formats as same in MPEG-4 format. The MP4 format was published in 2001. The first version of MP4 was revised and replaced into MPEG-4 Part14 in 2003 (ISO/IEC 14496-14:2003). ISO/IEC registered with codec version MJPEG, .Motion JPEG 2000, .MPEG-1, .MPEG-2 Part2, .MPEG-4 Part2/ASP, .Part10/AVC, .MPEG-H Part 2/HEVC [7].

AVI (Audio Video Interleave) introduced by Microsoft in 1992. It contains both Audio and video data files. AVI files create a file with no compression and resulting in large file size. It generally supports to seen with VLC Player. It does not provide standardized to encode, some approaches exist to support modern video compression techniques and functionality of MPEG4, although this intent of original specifications may cause problems [8].

MP4 or MPEG4 (Motion pictures expert group) was introduced for public use in 2001. It was developed by International organization of standard (ISO) includes .mp4, .m4p, .m4v, .m4r, m4a, .m4b. Different extensions of MP4 format introduced with different video codecs such H.262 / H.263 /H.264 / H.265 /MPEG-H where HEVC codecs implementation of the standard of video format (H.265) with an encoder (x265) to providing lower bitrate video compression files size and video quality.MPEG4 also known as Advanced Video Coding (AVC), mostly used in video coding formats is H.264 and AVC was introduced by MPEG in collaboration with International Telecommunication Union-Telecommunications (ITU-T) Video Coding Expert Group (VCEG). A well-known Microsoft family designs Microsoft video codecs know as MS MPEG-4v3 or DivX and WMV (Windows Media Video) include versions WMV7, WMV8, and WMV9. The latest standardized generation of WMV as the VC-1 standard [7] [9].

The TrueMotion introduced high definition (HD) video compression formats and codecs (VP6, VP6-E, VP6-S, VP7, VP8, and VP9) developed by (Duck cooperation) On2 Technologies used in such Adobe Flash player 8 and above versions and JavaFX and others desktop and mobile video platforms supports up to resolution 720P, 1080P and 2160P. The WebM (Web Movie) Format was introduced by Google launched in 2010. Google own self-use VP9 for YouTube service.VP8 and VP9 codecs available under the New BSD License by Google with source code. WebM supports audiovisual file formats that play on different search engines software such as Internet Explorer, Google Chrome, Mozilla Firefox, and Opera [10][11].

The High definition (HD) video streaming is still big problems for social cloud service providers for users with limited internet connection speed. HD video obviously contains big file size and storage space which proceeds additional time to upload on websites or web pages, it decreased the quality of service and needs more time in network bandwidth to transfer files cloud server to client side. Researchers put efforts to optimize Quality of Experience (QoE) and make perceptions for the end users. Studies in [12] presents video compression and other research in [13] explores issues of network bandwidth, nevertheless it is even more important to measure QoS, whereas QoS is more network-centric metric put efforts to make better QoS, however, PNSR (Peak-Signal to Noise Ratio) was used to quantify the QoS metrics (bandwidth, transmission, packet loss, delay) to measure the video quality. It included services of infrastructure, client, and network. These metrics are mostly concentrating on the quality of video and image.

A number of video quality measurements methods have been proposed [1]. QoS measurements different file format for video streaming on social clouds were never investigated by anyone in research work to analyze which file format provides better visual quality to end users with QoS. We conduct experiments and analysis of video streaming on different codecs and parameters over social clouds. During the experiments, we select different Video codec file formats which have different resolution, bitrate, frame rate and data rate. During the research, we repeat experiments with seven video file formats with different resolution quality (360P, 720P, 1080P and 4K), which were taken from YouTube. Downloaded videos were converted into 7 different standard video codecs and upload again to test on social clouds to examine video streaming and Quality of service which provides by social clouds.

This paper is organized into 5 sections. In section 2, we provide the literature review, section 3 is based on methodology, section 4 presents results and discussion; finally, we conclude our work in section 5.

## 2. Literature Review

In the past studies, the researcher presents a comparative study on different video file formats with different resolutions. Many researchers proposed model and schemes for the performance of video streaming to supports QoE but not QoS for Social clouds [1]. The proposed an algorithm works on different video format (FLV, MPEG4, and 3GP) using the multimedia parameters that determine the performance of video formats in the wireless network [5]. The frame rate of MPEG4 was examined and comparison, video standard and video sequences, HEVC monoscopic video codec was presented by Mallik et al [14] and proposed a model for low bitrate transmission with HEVC codec based on mixed resolution.

Research work provided by Laghari et al [15], Conducted QoE assessment on video file formats FLV, 3GP, WebM and MP4 with different resolutions of 240P, 360P, 720P and 2160P to quantify the user satisfaction about video quality. The work was based on user perception for different file formats and which one could better for online streaming, the results were given by users high rates for FLV 240P and WebM 360P for low-resolution videos whereas user assigned high MOS to MP4 with resolution 720P and 2160P.

Gohil et al [16] presented a comparative study of different video file formats and compression with minimum loss of data and quality. The proposed work was based on the conversion of original video file formats into other formats and different file size. Furthermore, [17] his proposed model works for advanced video compression technique which was developed on the base of the previous survey. This model functioned on the limited version for making the video more efficient than MPEG4 and H264 and improves the video quality and reduces the file size of storage without decreasing the quality to transfer over the network.

Kamadyta et al. [18] analyzed the performance of a wireless network which supports multimedia transmission in IEEE 802.11e. They used three distinct types of traffic (voice, Video, and data). This work used the EDCA method to differentiate IEEE 802.11 DCF and IEEE 802.11e EDCA with CFB scheme for reducing delay and improve throughputs.

Lin et al [19] proposed a work for QoS guarantees for wireless scalable video streaming and cross-layer optimization MCS (Modulation and Coding Scheme) for HD video streaming (Scalable) considered in video rate, distortion parameter, terminal or channel conditions, transmission duration and QoS metric. This scheme minimized the video distortion of video streaming and the duration of the transmission period by selected and targeted video. The video rate and payload length of packets under given delay bound.

In the past, only QoE user perceptions was considered for the comparative study of file formats with different resolution and QoS of video streaming is analyzed on artificial network parameters [20] but QoS was never be monitored for social clouds, network parameters, and compression video parameters. We preferred to use networking website of social clouds such as Facebook, Google+ in our experiments to analyze results.

## 3. Design and Experiment (Methodology)

In order to assess the QoS of video streaming, we performed various experiments by using different videos on social clouds. Technical work based on HD videos with four different resolution and quality, 360P, 720P, 1080P and 4K, which were selected from YouTube. Downloaded videos were converted through software Format Factory 4.3.0 [21] into standardized video file formats such as (MOV, FLV, WebM, WMV, AVI, MP4 (HEVC and AVC). These four different resolution videos were combined in 7 standard video file formats tests were made for each combination, making it 28 experiment tests in total. Videos were played in sequence by standard YouTube video streamer which is compatible to use on desktop and mobile device. During the posting of videos, cloud reduces the quality of video (data rate, bitrate, and size) which varies secret of compression on cloud preference.

Standardized and original videos were uploaded on social clouds and then played from the hosted cloud in real time streaming. The experimental methodology is given in figure 1.

Before downloading and playing the online videos, network speed was measured by popular speedtest site [22]. The upload speed of the network was 3.50Mb/s and downloads were 10.09 Mb/s during the experiment. The purpose of selecting the low-quality video (360P) to high (4k=2160P) resolution to assess the QoS for video streaming experiment on low quality and high quality. The technical details of videos such as video file format, file size, data rate, bitrate, video stream size and duration are given in Table1.

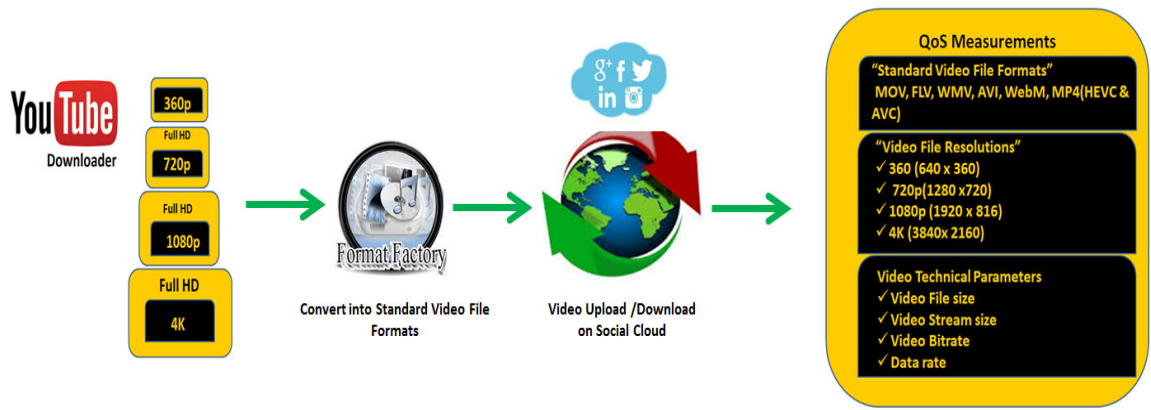

Fig 1. Experimental Methodology

Table 1. Original Video File Formats and Parameters

| Video File Format-360P | | File Size | bitrate | Data rate | Video Stream Size | Duration |
|---|---|---|---|---|---|---|
| MOV Format | HEVC 360 | 2.04M | 274 kb/s | 135kb/s | 1.00 MB (49%) | |
| FLV Format | FLV1 360 | 4.60M | 617 kb/s | - | 3.72 MB (81%) | |
| WMV Format | 360 | 7.05M | 274 kb/s | 768kb/s | 5.72 MB (81%) | |
| AVI Format | 360 | 5.07 M | 680 kb/s | 512kb/s | 4.02 MB (79%) | 1 min 2 sec |
| WEBM Format | 360 | 1.83 M | 246 kb/s | - | - | |
| MP4 Format | HEVC 360 | 2.04 M | 274 kb/s | 135kb/s | 1.00 MB (49%) | |
| MP4 Format | AVC 360 | 3.75 M | 502 kb/s | 366kb/s | 2.71 MB (72%) | |
| Video File Format-720P | | File Size | bitrate | Data rate | Video Stream Size | Duration |
| MOV Format | HEVC | 9.32 M | 2601 kb/s | 2596kb/s | 9.28 MB (100%) | 30s 70 ms |
| FLV Format | FLV 720P | 8.28 M | 2309 kb/s | - | 7.17 MB (87%) | 30s 93 ms |
| WMV Format | 720i | 3.95M | 1103 kb/s | 887kb/s | 3.30 MB (84%) | 30s 76 ms |
| AVI Format | DIV x 720P | 8.03 M | 2238 kb/s | 2048kb/s | 7.51 MB (94%) | 30s 93 ms |
| WEBM Format | 720P | 1.13 M | 316 kb/s | - | - | 30s 54 ms |
| MP4 Format | HEVC 720 | 9.33 M | 2602 kb/s | 2596kb/s | 9.28 MB (100%) | 30s 68 ms |
| MP4 Format | AVC 720i | 5.68 M | 1586 kb/s | 1578kb/s | 5.65 MB (99%) | 30s 70 ms |
| Video File Format-1080P | | File Size | bitrate | Data rate | Video Stream Size | Duration |
| MOV Format | HEVC 1080P | 28.29M | 7697 kb/s | 7708kb/s | 27.6 MB (97%) | 30s 835 ms |
| FLV Format | FLV1 1080P | 27.26M | 7412 kb/s | - | 14.7 MB (54%) | 30s 850 ms |
| WMV Format | WMV2 | 10.99M | 2991 kb/s | 2917kb/s | 10.1 MB (91%) | 30s 835 ms |
| AVI Format | DIVX | 11.39M | 3097 kb/s | 2979kb/s | 10.9 MB (95%) | 30s 851 ms |
| WEBM Format | VP8 1080P | 4.72M | 1284 kb/s | - | - | 30s 820 ms |
| MP4 Format | HEVC 1080P | 28.29M | 7696 kb/s | 7707kb/s | 27.6 MB (97%) | 30s 835 ms |
| MP4 Format | AVC 1080P | 10.70M | 2910 kb/s | 2853kb/s | 10.2 MB (95%) | 30s 837 ms |
| Video File Format-4k | | File Size | bitrate | Data rate | Video Stream Size | Duration |
| MOV Format | HEVC 3840i | 11.78M | 9800 kb/s | 9878kb/s | 11.8 MB (100%) | 10s 84 ms |
| FLV Format | FLV 3840i | 9.28M | 7720 kb/s | - | 8.72 MB (94%) | 10s 83 ms |
| WMV Format | WMV2 3840i | 11.97M | 9955 kb/s | 9675kb/s | 11.6 MB (97%) | 10s 84 ms |
| AVI Format | MPEG4 DIVX3840i | 11.83M | 9839 kb/s | 9675kb/s | 11.8 MB (100%) | 10s 83 ms |
| WEBM Format | VP8 3840i | 1.81M | 1505 kb/s | - | 1.73 MB (96%) | 10s 84 ms |
| MP4 Format | HEVC 3840i | 11.8 MB | 9800 kb/s | 9878kb/s | 11.8 MB (100%) | 10s 84 ms |
| MP4 Format | AVC 3840i | 9.28M | 7718 kb/s | 7780kb/s | 9.27 MB (100%) | 10s 84 ms |

## 4. Result and Discussion

During the experiments, File viewer lite and VLC (version 3.0.2) players were used for playing video [23, 24], which were downloaded from social clouds. All video parameters were analyzed and generate log file reports by using MediaInfo software tool (version 8.02) [25]. The functionality of Media-Info tool is to decodes all information of videos, which were compressed videos during upload and download process. Table 2 provides information of 360P video file format of each cloud compared to original video file. The videos have a difference in file size, data rate and bitrate. Technical parameters such as data rate and bitrate can help to know about streaming of the video quality either it has a large file, good and worst video quality in future and supports in minimize bandwidth requirements for media transmission.

### 4.1 Standard Video File Formats (360P)

Low quality (360P) standard video File Format with (640 x 360) frame height and width, 24frames/sec frame rate videos were uploaded/ downloaded on the social cloud to find the differences of QoS and quality. During play online video from the social cloud, network parameters were not analyzed because network speed was support video streaming without buffering and delay situation. By analyzing at the technical parameters of the video were measured in file size, data rate, total bitrate, video stream size and duration of playing length respectively.

QoS Impact of video streaming is different on the cloud as compared to original video. Facebook compressed videos by decreasing data rate and bitrate to reduce the video file size on the upload. For example, MOV reduced 4.9% storage file size down to 1.94MB as the original was 2.04 MB but does not have a high impact on video streaming and quality. The same way FLV format is compressed in terms of data rate, and bitrate to 53% which reduced the file size 2.66 MB as low and original was 4.60 MB. For WMV file format decreased 44.8% video bitrate, 54.5% data rate to reduce storage size 2.97MB as the original was 7.05MB. AVI slightly better compressed to 46.7% bitrate, 38.8% data rate and 46.7% reduce storage file size 2.70MB as compared to original 5.07MB. WebM increased 12% storage size and 10% bitrate; MP4 (HEVC) increased 26.6% video bitrate, 61.3% data rate and 45.5% storage size and MP4 (AVC) decreased 23.5% video bitrate, 31.9% data rate to reduce 23.2% storage file size.

The same way followed in Google+ cloud, which slightly less compressed the video bitrate and data rate but did not compress the storage file size. In MOV decreased 1.9% video bitrate, WMV increased 69.4% video bitrate; AVI reduced 5.8% video bitrate, WebM decreased 0.5% storage file size , MP4 (HEVC) increased 1.04% in storage file size 3.74MB , 45.2% video bitrate and 35.8% data rate, and  MP4 (AVC) decreased 45.6% storage size, 46.4% video bitrate and 63.1% data rate as low as 2.04 MB.

Google+ provided good quality for downloaded video conversely the QoS are better than other clouds to be considered same as it in the video file formats without exchanging techniques and methods on video file formats but the online played video of Google+ required more network bandwidth; situation happened in buffering /delay due to low network speed. Effects of video streaming on quality during posting and during the online streaming from social and technical parameters comparison of the original video with posted videos on social clouds are given in Table2.

Table 2. Comparison of original with cloud compressed video parameters (video File format 360P)

| File Format | File size | | | Video bitrate | | | Video Data rate | | |
|---|---|---|---|---|---|---|---|---|---|
| 360P | Original | Facebook | Google+ | Original | Facebook | Google+ | Original | Facebook | Google+ |
| MOV | 2.04M | 1.94 MB | 2.04 MB | 274 kb/s | 257kb/s | 269kb/s | 135kb/s | 208kb/s | 135kb/s |
| FLV | 4.60M | 2.66 MB | 4.60 MB | 617 kb/s | 357kb/s | - | - | 308kb/s | - |
| WMV | 7.05M | 2.97 MB | 7.05 MB | 274 kb/s | 397kb/s | 896kb/s | 768kb/s | 349kb/s | 768kbpd |
| AVI | 5.07 M | 2.70 MB | 5.07 MB | 680 kb/s | 362kb/s | 640kb/s | 512kb/s | 313kb/s | 512kb/s |
| WEBM | 1.83 M | 2.05 MB | 1.82 MB | 246 kb/s | 272kb/s | - | - | 224kb/s | - |
| MP4(HEVC) | 2.04 M | 2.97 MB | 3.74 MB | 274 kb/s | 347kb/s | 500kb/s | 135kb/s | 349kb/s | 366kb/s |
| MP4 (AVC) | 3.75 M | 2.88 MB | 2.04 MB | 502 kb/s | 384kb/s | 269kb/s | 366kb/s | 249kb/s | 135kpbs |

## 4.2 Standard Video File Formats (720P)

The technical parameters of 720P video format with (1280 x720) frame height width and 29 frames/sec frame rate was uploaded on the social clouds and parameters of the video were changed as compared to low-quality 360p by applying compression techniques by social clouds. 720P required faster network as compared to 360P due to high video quality and file size. Facebook compressed video by decreasing video bitrate; data rate and storage file size. For example, MOV compressed 72.6% video bitrate, 74.4% data rate to reduce 72.5% file size in 2.56MB the original one was 9.32 MB. FLV compressed 44.9% video bitrate, 80.4% data rate, 46.7% reduced to storage file size, WMV decreased 71.7% video bitrate, 70.2% data rate and 24.6% compressed the file size, AVI  reduced 51.2% file size , 40.6% video bitrate and 37.5% data rate. WebM increased the file size 2.70MB as compared to original video File size 1.13MB, MP4 (HEVC) reduced 72.6% video bitrate, 74.4% data rate and compressed 72.5% file size as low as 2.56MB, the original one was 9.33MB and MP4 (AVC) format slightly compressed in 50.7% video bitrate, 50.6% data rate and 43.9% to reduced file size in 2.82MB as compared 5.68 MB respectively.

The same way videos were uploaded on Google+ social cloud, it compressed only video bitrate and data rate but does not effect on Video file format and storage file size, for example, MOV format decreased 0.1% video bitrate but did not compressed data rate and storage file size. Technical parameters of FLV file format did not compress. Video bitrate of WMV file was compressed 7.9%. AVI decreased 2.7% video bitrate, WebM, MP4 (HEVC) decreased 0.1% of video bitrate and increased 0.07% data rate. MP4 (AVC) decreased 0.3% video bitrate respectively. The effects of video quality and technical parameters comparison of original video posted videos on the clouds are given in Table 3.

Table 3. Comparison of original with cloud compressed video parameters (video File format 720P)

| File Format | Video File size | | | Video bitrate | | | Video Data rate | | |
|---|---|---|---|---|---|---|---|---|---|
| 720P | Original | Facebook | Google+ | Original | Facebook | Google+ | Original | Facebook | Google+ |
| MOV | 9.32 M | 2.56MB | 9.32 M | 2601 kb/s | 712kb/s | 2598kb/s | 2596kb/s | 663kb/s | 2596kb/s |
| FLV | 8.28 M | 4.56 MB | 8.28 M | 2309 kb/s | 1271kb/s | 2309kb/s | 821kb/s | 1222kb/s | 821kb/s |
| WMV | 3.95M | 1.13 MB | 3.95M | 1103 kb/s | 312kb/s | 1015kb/s | 887kb/s | 264kb/s | 887kb/s |
| AVI | 8.03 M | 4.76 MB | 8.03 M | 2238 kb/s | 1328kb/s | 2176kb/s | 2048kb/s | 1279kb/s | 2048kb/s |
| WEBM | 1.13 M | 2.70 MB | 1.13 M | 316 kb/s | 751kb/s | - | - | 703kb/s | - |
| MP4(HEVC) | 9.33 M | 2.56 MB | 9.33 M | 2602 kb/s | 712kb/s | 2598kb/s | 2596kb/s | 663kb/s | 2598kb/s |
| MP4 (AVC) | 5.68 M | 2.82 MB | 5.68 M | 1586 kb/s | 781kb/s | 1580kb/s | 1578kb/s | 779kb/s | 1578kb/s |

## 4.3 Standard Video File Formats (1080P)

The same way was applied for Video File Formats (1080P) with (1920 x 816) frame width .height; 29 fps frame rate uploaded on the social cloud. The 1080P required faster network speed to upload on the cloud as compared to other standard video 360P and 720P for video streaming. For example, Facebook compressed MOV parameters 83.7% video bitrate, 84.3% data rate to reduce the 12.2% file size as compared to original size 28.29MB and Google+ compressed MOV parameters where 89.7% video bitrate but does not effect on file size and data rate. Facebook changed parameters of FLV as well to reduced 7.7% file size, 80.6% bitrate.WMV decreased 39.2% storage file size, 50.2% video bitrate, 50.5% data rate. AVI decreased 51% storage file size and decreased 47.5% video bitrate and 51.1% data rate by Facebook and Google+ increased

1.8% video bitrate. WebM format has a small file size as compare to other video file sizes because it has 1284kb/s original video bitrate then Facebook compressed 29.1% 910kb/s bitrate. MP4(HEVC) has big file size 28.29MB, which compressed in 84.3% into new file size 4.44MB, 87.85% video bitrate, 84.6% data rate by Facebook and Google+ reduced 62.5% storage file size,61.2% video bitrate and 62.9% data rate but does not impact on video quality. Facebook in MP4 (AVC) reduced 57.9% storage file size, 56.9% video bitrate, 57.83% data rate and Google+ increased the storage file 28.2MB as the original one was 10.70MB, it has maximized the 62% file size , 63.1% video bitrate and 62.9% data rate, it  has no effects on video quality . The technical parameters comparison of the original video with posted videos on social clouds is given in Table 4.

Table 4. Comparison of original with cloud compressed video parameters (video File format 1080P)

| File Format | Video File size | | | Video bitrate | | | Video Data rate | | |
|---|---|---|---|---|---|---|---|---|---|
| 1080P | Original | Facebook | Google+ | Original | Facebook | Google+ | Original | Facebook | Google+ |
| MOV | 28.29M | 4.50MB | 28.2 MB | 7697 kb/s | 1252kb/s | 790kb/s | 7708kb/s | 1203kb/s | 7708kb/s |
| FLV | 27.26M | 5.31MB | 27.2 MB | 7412 kb/s | 1431kb/s | - | - | 1383kb/s | - |
| WMV | 10.99M | 5.52MB | 10.9 MB | 2991 kb/s | 1489kb/s | 3045kb/s | 2917kb/s | 1441kb/s | 2917kb/s |
| AVI | 11.39M | 5.58MB | 11.3 MB | 3097 kb/s | 1504kb/s | 3107kb/s | 2979kb/s | 1456kb/s | 2979kb/s |
| WEBM | 4.72M | 3.27MB | 4.71 MB | 1284 kb/s | 910kb/s | - | - | 860kb/s | - |
| MP4(HEVC) | 28.29M | 4.44MB | 10.6 MB | 7696 kb/s | 1235kb/s | 2982kb/s | 7707kb/s | 1186kb/s | 2853kb/s |
| MP4 (AVC) | 10.70M | 4.50MB | 28.2 MB | 2910 kb/s | 1253kb/s | 7901kb/s | 2853kb/s | 1203kb/s | 7707kb/s |

## 4.4 Standard Video File Formats (4K=2160p)

High quality Video (4k=2160P) with (3840x 2160) resolution, 24 frames/sec frame rate  was uploaded on social cloud and comparison of compressed results shows the Facebook compressed the original MOV file 11.78MB to 859 KB , to reduced 92% video bitrate, 92.9% data rate, FLV compressed 89.7% file size, 89.9% bitrate.WMV compressed 85.7% storage file size , 85.6% bitrate, 85.1% data rate.WMV reduced 85.7% file size, 85.6% bitrate, 85.1% data rate. AVI decreased 87.9% file size, 87.8% bitrate and 87.6% data rate. WebM compressed original 1.81MB file size into 925KB at  48.8% and 49.8% bitrate.MP4(HEVC)  decreased 92.7% file size, 92.8% bitrate size and 92.9% data rate. MP4 (AVC) original file size was 9.28MB; it was compressed 854 KB at 90.7%, reduced 8% bitrate and 91% data rate.

Google+ increased MOV parameters in 0.7% video bitrate but did not compressed data rate and storage file size. FLV

compressed 0.1% file size. WMV increased 2.8% bitrate, AVI decreased 1.6% bitrate, MP4 (HEVC) increased 0.7% bitrate and MP4(AVC) increased 0.8% video bitrate. Effects of video quality are better than 1080P video file formats. The technical parameters comparison of the original video with posted videos on social clouds is given in Table 5.

Table 5. Comparison of original with cloud compressed video parameters (video File format 4K=2160p)

| File Format | Video File size | | | Video bitrate | | | Video Data rate | | |
|---|---|---|---|---|---|---|---|---|---|
| (4K) | Original | Facebook | Google+ | Original | Facebook | Google+ | Original | Facebook | Google+ |
| MOV | 11.78M | 859KB | 11.7 MB | 9800 kb/s | 701kb/s | 9878kb/s | 9878kb/s | 701kb/s | 9878kb/s |
| FLV | 9.28M | 953KB | 9.27 MB | 7720 kb/s | 777kb/s | - | - | 777kb/s | - |
| WMV | 11.97M | 1.71MB | 11.9 MB | 9955 kb/s | 1432kb/s | 9675kb/s | 9675kb/s | 1432kb/s | 9675kb/s |
| AVI | 11.83M | 1.42MB | 11.8 MB | 9839 kb/s | 1195kb/s | 9675kb/s | 9675kb/s | 1195kb/s | 9675kb/s |
| WEBM | 1.81M | 925KB | 1.80 MB | 1505 kb/s | 755kb/s | - | - | 755kb/s | - |
| MP4(HEVC) | 11.8 M | 859KB | 11.7 MB | 9800 kb/s | 701kb/s | 9878kb/s | 9878kb/s | 701kb/s | 9878kb/s |
| MP4 (AVC) | 9.28M | 854KB | 9.27 MB | 7718 kb/s | 697kb/s | 7780kb/s | 7780kb/s | 697kb/s | 7780kb/s |

## 5. Comparison of Video Streaming Size

Videos were loaded on social clouds web pages in different resolution 360P, 720P, 1080P and 4K. During the process, the videos were streamed depending on the secrets of compression of social clouds. Some videos actually stream at lower to higher resolution and size and tend to increase or decrease bitrate and data rate size after a video uploaded/ downloaded with encoding process on the social clouds. Effects of video quality and technical comparison of the original video stream size with posted videos stream size on social clouds are given in Figure 2.

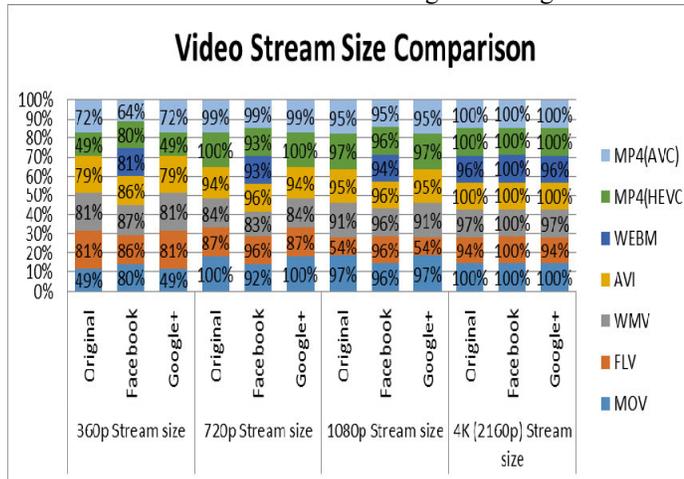

Fig 2. Comparison of the original video stream size with posted stream size

## 6. Conclusions

In this paper, we conducted experiments on social clouds to upload 7 different HD video file formats such as MOV, FLV, WMV, AVI, WebM, MP4( HEVC and AVC) having different resolutions of 360P, 720P, 1080P and (4k=2160P) respectively to measure the technical parameters for video streaming. Standard video codecs supported streaming platform which is mostly used for online streamed from the web but the video is automatically streamed when the page is loaded. The results summarized in this paper that low-resolution 360P compressed low bitrate and data rate as compared Full HD 720P, 1080P and 2160P which required high compression of file size and required high network speed to transfer a video file from the client to cloud or cloud to the client. The results included Facebook reflects all video formats changed into MP4 formats so furthermore in details this paper shown the Facebook provides good visual quality for HD videos after video compression on different scales. Examples are (4K =2160P) Provides 100% of video streaming in MP4 (AVC) codec and provide the best quality on Facebook and enhances the QoE on the low bandwidth of networks as compared to Google+, which provide best QoS. WebM file format is newly introduced by Google which compressed more and provides same video quality as compared to MP4 and their all technical parameters like storage size, a total bitrate that required low network bandwidth for video streaming which also saves timescale and resources for transmission.

We believe this research work has a great potential to explore cloud services related issues and improve the overall the QoS. On the other hand, this will beneficial for the whole community of cloud networks and operators and service providers to use particular video file format and codec with resolution and size according to the availability of network speed.

**Acknowledgment**
The work is supported by the National Key R&D Program of China under grant No.2017YFB0801801, the National Science Foundation of China (NSFC) under grant No. 61472108 and 61672186..

**Competing Interests**
Declare Conflicts of interest of state "The authors declare no conflicts of interest".

## Biographies of Authors

**Sajida Karim** received the B.S. and Master's degree in Computer science and Information Technology from Shah Abdul Latif University, Khairpur, Pakistan, in 2017 respectively; From 2013-2017, she was a Lecturer in the Computer science and Information Technology Department, Shah Abdul Latif University. She is currently pursuing a Ph.D. degree with the Harbin Institute of Technology HIT, Harbin China. She published technical articles in scientific journals and conference proceedings. Her current research interests include Computer networks, Cloud Computing, multimedia QoS/QoE management, Artificial Intelligence, Deep Learning Neural Networks.

**He Hui** is Dr., associate professor, member of IEEE & IEEE Computer , China Computer Society, ACM Association. Harbin Institute of Technology School of computer science and technology. Mainly engaged in computer network, network measurement and simulation, network active defense technology, mobile network security, cloud computing, migration learning and so on. Presided over or participated in the national network information security project key projects Natural Science Foundation, 863 projects, 973 projects and other provincial ministries and commissions more than 10 items, won the national defense science and technology progress prize two, Heilongjiang province science and technology two prize 1 item. 2010 to guide students to participate in the national information security competition won the first prize 1. At home and abroad important academic journals and conferences published nearly 80 papers..

**Asif Ali Laghari** received the B.S. degree in Information from the Quaid-eAwam University of Engineering Science and Technology Nawabshah, Pakistan, in 2007 and Master degree in Information Technology from the QUEST Nawabshah Pakistan in 2014.From 2007 to 2008, he was a Lecturer in the Computer and Information Science Department, Digital Institute of Information Technology, Pakistan,. He received PhD degree from the school of the Computer Science & Technology, Harbin Institute of Technology. He has published technical articles in scientific journals and conference proceedings. His current research interests include Computer networks, cloud computing, and multimedia QoE management.

**Hina Madiha** received the B.E. degree in Computer Systems Engineering from Quaid-e-Awam University of Engineering, Science & Technology Nawabshah, Pakistan in 2016. She is currently enrolled in Master degree program in School of Computer Science Shaanxi Normal University Xi'an, China. Her research interests include cloud computing, image processing, Artificial Intelligence, and Machine learning.

IJCSI
www.IJCSI.org